\documentclass{article}
\usepackage{arxiv}
\usepackage{amsmath}  
\usepackage{fancyhdr} 

\usepackage[utf8]{inputenc} 
\usepackage[T1]{fontenc}    
\usepackage{hyperref}       
\usepackage{url}            
\usepackage{booktabs}       
\usepackage{amsfonts}       
\usepackage{nicefrac}       
\usepackage{microtype}      
\usepackage{lipsum}		
\usepackage{graphicx}
\usepackage{natbib}
\usepackage{doi}

\renewcommand{\headeright}{Technical Report}
\renewcommand{\undertitle}{Technical Report}
\renewcommand{\shorttitle}{Fish-Speech}

\title{Fish-Speech: Leveraging Large Language Models for Advanced Multilingual Text-to-Speech Synthesis}
\author{
    Shijia Liao\textsuperscript{1} \and
    Yuxuan Wang\textsuperscript{1} \and
    Tianyu Li\textsuperscript{1} \and
    Yifan Cheng\textsuperscript{1} \and
    Ruoyi Zhang\textsuperscript{1} \and
    Rongzhi Zhou\textsuperscript{1} \and
    Yijin Xing\textsuperscript{1} \\[2ex]
    \textsuperscript{1}Fish Audio \\[2ex]
    \texttt{\{lengyue,honst,stardust\}@fish.audio, yf\_cheng@hust.edu.cn,} \\
    \texttt{potato\_zhang@nuist.edu.cn, anya\_zhou@bjtu.edu.cn, rcell233@outlook.com}
}
\date{}

\hypersetup{
pdftitle={Fish-Speech: Leveraging Large Language Models for Advanced Multilingual Text-to-Speech Synthesis},
pdfsubject={Fish-Speech: Leveraging Large Language Models for Advanced Multilingual Text-to-Speech Synthesis},
pdfauthor={Shijia Liao, ~Tianyu Li, ~Yifan Cheng, ~Ruoyi Zhang, ~Rongzhi Zhou, ~Yuxuan Wang, ~Rcell},
pdfkeywords={Fish Speech, Text-to-Speech},
}
    
\begin{document}

\maketitle

\begin{abstract}
Text-to-Speech (TTS) systems face ongoing challenges in processing complex linguistic features, handling polyphonic expressions, and producing natural-sounding multilingual speech - capabilities that are crucial for future AI applications. In this paper, we present Fish-Speech, a novel framework that implements a serial fast-slow Dual Autoregressive (Dual-AR) architecture to enhance the stability of Grouped Finite Scalar Vector Quantization (GFSQ) in sequence generation tasks. This architecture improves codebook processing efficiency while maintaining high-fidelity outputs, making it particularly effective for AI interactions and voice cloning.

Fish-Speech leverages Large Language Models (LLMs) for linguistic feature extraction, eliminating the need for traditional grapheme-to-phoneme (G2P) conversion and thereby streamlining the synthesis pipeline and enhancing multilingual support. Additionally, we developed FF-GAN through GFSQ to achieve superior compression ratios and near 100\% codebook utilization.

Our approach addresses key limitations of current TTS systems while providing a foundation for more sophisticated, context-aware speech synthesis. Experimental results show that Fish-Speech significantly outperforms baseline models in handling complex linguistic scenarios and voice cloning tasks, demonstrating its potential to advance TTS technology in AI applications. The implementation is open source at \href{https://github.com/fishaudio/fish-speech}{https://github.com/fishaudio/fish-speech}.
\end{abstract}


\section{Introduction}

The past decade has seen remarkable progress in Text-to-Speech (TTS) systems, transforming applications from virtual assistants to educational tools. Current TTS architectures, such as VALL-E [\cite{wang2023neural}], VITS [\cite{kim2021conditional}], FastSpeech [\cite{ren2020fastspeech}] typically rely on grapheme-to-phoneme (G2P) conversion [\cite{klatt1987review}] to convert text into phonetic representations before synthesis. While effective, this approach struggles with context-dependent polyphonic words and cross-lingual generalization due to complex phonetic rules. Recent advances in zero-shot voice conversion such as YourTTS [\cite{casanova2022yourtts}] and unified speech generation model UniAudio [\cite{yang2023uniaudio}] have shown the potential of neural architectures in handling various speech tasks. Additionally, flow-based models like CosyVoice[\cite{du2024cosyvoice}, MatchaTTS\cite{mehta2024matcha}] have demonstrated promising results in natural speech synthesis. However, most solutions disentangled semantic and acoustic feature as a trade-off to improve stability, and reduce the voice cloning understanding ability.

As demand grows for multilingual TTS systems, the limitations of G2P-based approaches become more apparent. The need for language-specific phonetic rules and lexicons hinders scalability and complicates system maintenance. Recent research has explored the use of Large Language Models (LLMs) for direct linguistic feature extraction, eliminating the need for explicit G2P conversion. [\cite{betker2023better}].

We introduce Fish-Speech, a novel TTS framework featuring a serial fast-slow dual autoregressive (Dual-AR) architecture. This design improves the stability of grouped finite scalar vector quantization (GFSQ) in sequence generation while maintaining high-quality output. By incorporating LLMs into the TTS pipeline, Fish-Speech simplifies the synthesis process and better handles polyphonic characters and multilingual text. The model trains on 720,000 hours of multilingual audio data, enabling it to learn diverse linguistic patterns and pronunciation variations.

To improve synthesis quality, we develop Firefly-GAN (FFGAN), a new vocoder architecture based on Grouped Finite Scalar Vector Quantization(GFSQ). FFGAN combines Finite Scalar Quantization (FSQ) [\cite{mentzer2023finite}], and Group Vector Quantization (GVQ) to optimize compression ratios and codebook usage. Our evaluations show 100\% codebook utilization, representing state-of-the-art performance in this field.

The primary contributions of this work are as follows:
\begin{itemize}
    \item We introduce Fish-Speech, a novel TTS framework that leverages LLMs and a Dual-AR architecture to replace traditional G2P conversion, providing robust and scalable multilingual speech synthesis.
    \item We present FFGAN, an advanced vocoder that integrates multiple vector quantization techniques to achieve high-fidelity speech synthesis with optimized compression ratios and codebook utilization.
    \item We develop fish-tech acceleration methodologies, the system achieves real-time factors of approximately 1:5 on consumer-grade NVIDIA RTX 4060 mobile platforms and 1:15 on high-performance NVIDIA RTX 4090 configurations. And a latency of 150ms which is far less than other TTS system using DiT and Flow structure.
\end{itemize}

We encourage readers to listen to our samples on \href{https://speech.fish.audio/samples/}{fish speech 1.4 sample}. We also highly recommend that you go to our online synthesis site \href{https://fish.audio}{fish.audio} to try out the different speakers of audio synthesized by the community.

\section{Related Work}
\setcounter{subsection}{0}

\subsection{Text-to-Speech Systems}

Text-to-Speech (TTS) systems have evolved dramatically from basic phoneme-based models to sophisticated end-to-end neural approaches that directly convert text to speech [\cite{tan2021survey}]. This transformation, driven by advances in deep learning and increased computational power, has led to major improvements in speech naturalness, prosody control, and cross-language capability [\cite{ren2019almost}]. Modern TTS systems now serve diverse applications, from intelligent assistants to accessibility tools and human-computer interfaces [\cite{capes2017siri}].

\subsection{Neural Vocoders}

Neural vocoders have played a key role in improving speech synthesis quality. WaveNet [\cite{van2016wavenet}] first introduced autoregressive models for audio generation, followed by more efficient architectures like WaveRNN [\cite{kalchbrenner2018efficient}] and WaveGrad [\cite{chen2020wavegrad}]. HiFi-GAN [\cite{kong2020hifi}] later introduced adversarial training, setting new standards in audio quality and computational efficiency. EVA-GAN is a brand new GAN structure vocoder developed by NVIDIA [\cite{liao2024eva}], it uses a Context Aware Module(CAM) for improved performance with minimal computational overhead. EVA-GAN shows superior performance over existing state-of-the-art vocoders in both objective and subjective metrics, particularly in spectral continuity and high-frequency reconstruction. 

\subsection{Vector Quantization in Speech Synthesis}

Vector Quantization (VQ) has become essential in modern speech synthesis. VQ-VAE [\cite{van2017neural}] showed the effectiveness of discrete latent representations for audio generation, while SoundStream [\cite{zeghidour2021soundstream}] and EnCodec [\cite{defossez2022high}] further improved these techniques for high-quality audio compression and synthesis.

\subsection{Large Language Models in Speech Processing}

Large Language Models (LLMs) are increasingly important in speech processing. Nowadays there are more and more models using BERT, huBERT as an intermediate structure for TTS, such as Parler TTS [\cite{lacombe-etal-2024-parler-tts}], Melo TTS [\cite{zhao2024melo}], E3-TTS [\cite{gao2023e3}], XTTS [\cite{casanova2024xtts}] and so on. They all achieve better synthesis effect.

\subsection{Multilingual Speech Synthesis}

Multilingual speech synthesis faces unique challenges in maintaining consistent quality across languages. Recent solutions include unified multilingual models [\cite{liu2019cross}], cross-lingual transfer learning [\cite{nekvinda2020one}], and language-agnostic representations [\cite{li2019bytes}].

\section{Methods}

Fish-Speech is a novel Text-to-Speech (TTS) framework that addresses key limitations of current non-grapheme-to-phoneme (non-G2P) TTS systems. The framework is specifically designed to handle multi-emotional and multilingual speech synthesis, with a focus on meeting the demands of advanced AI conversational agents.

Building on recent advances in vector quantization and condition representation [\cite{kumar2024high}, \cite{chen2023vector}, \cite{wang2019vector}], we introduce a Grouped Finite Scalar Vector Quantization(GFSQ) technique. This method efficiently encodes latent conditions, enabling better capture and reproduction of subtle speech variations. Our approach achieves 100\% codebook utilization, maximizing the effectiveness of the quantization space.

We also develop a dual autoregressive (dual-AR) architecture that solves two major challenges in current TTS systems. First, it improves the stability of code generation, a common issue in existing frameworks. Second, it offers better generation efficiency compared to Diffusion Transformers (DiT), making it well-suited for real-time applications. Last, but most importantly, it is ready for Voice Agent, which we will release in near future.

\begin{figure}[h]
\vspace{5pt}
\centering
\includegraphics[width=0.6\textwidth]{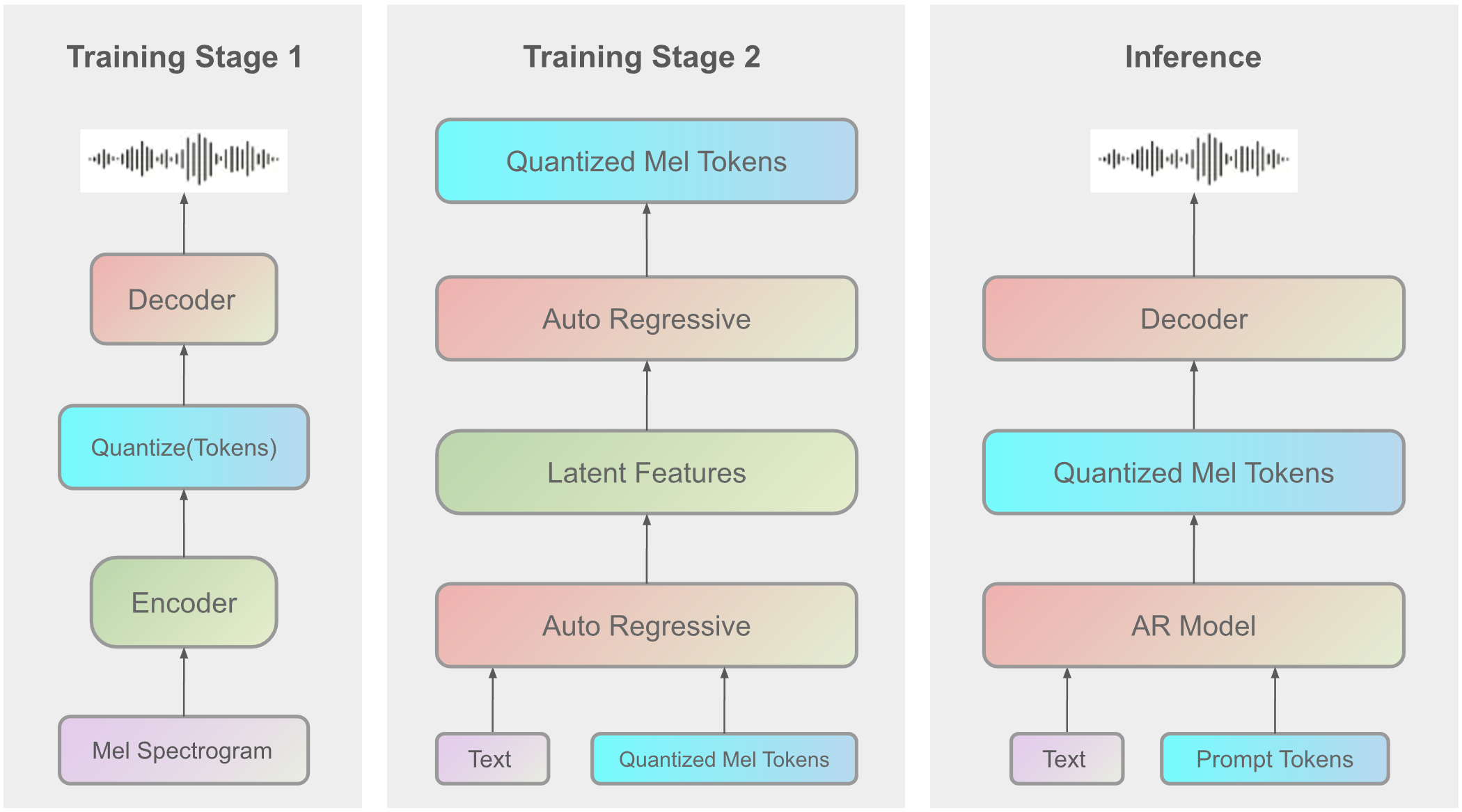}
\caption{Fish Speech Architecture}
\label{fig:fs_architecture}
\end{figure}

\subsection{Dual Autoregressive Architecture in Fish-Speech}

\begin{figure}[h]
\vspace{-23pt}
\centering
\includegraphics[width=1\textwidth]{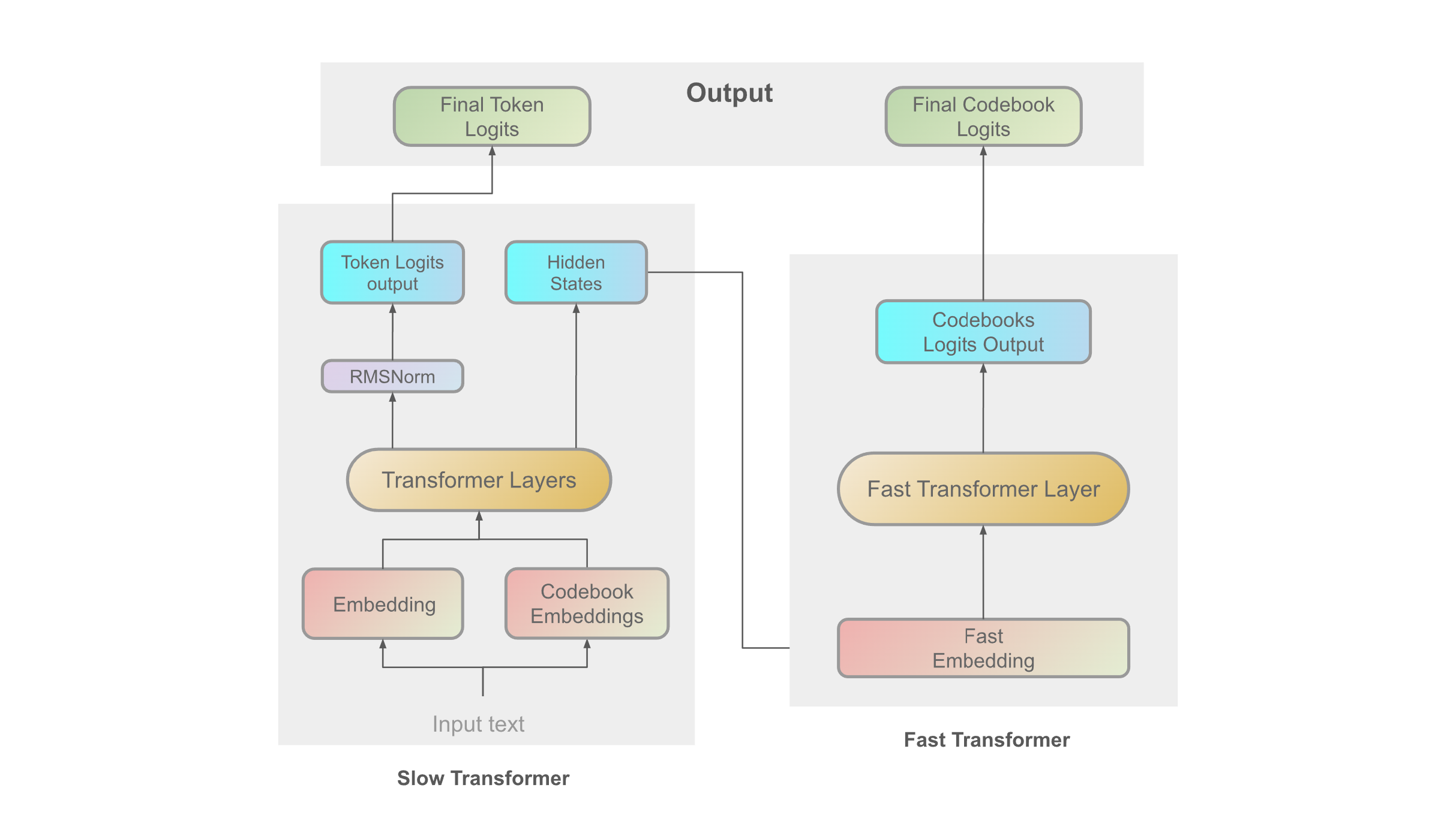}
\caption{Architectural overview of the Dual Autoregressive (Dual-AR) framework in Fish-Speech.}
\label{fig:dual-ar-overview}
\end{figure}

This section describes the \textbf{Dual Autoregressive (Dual-AR) architecture} [Fig.~\ref{fig:dual-ar-overview}] of Fish-Speech, a TTS system designed to handle complex linguistic features, polyphonic words, and natural-sounding multilingual synthesis. The Dual-AR architecture improves the stability and computational efficiency of codebook processing during sequence generation, particularly when using Grouped Finite Scalar Vector Quantization (GFSQ).

\subsubsection{Overview of the Dual-AR Architecture}

The Dual-AR architecture consists of two sequential autoregressive transformer [\cite{vaswani2017attention}, \cite{subakan2021attention}] modules: a \textbf{Slow Transformer} and a \textbf{Fast Transformer} [\cite{yang2023uniaudio}]. This design processes both high-level and detailed aspects of speech synthesis efficiently.
\textbf{\subsubsubsection{Slow Transformer}}

The Slow Transformer processes input text embeddings to capture global linguistic structures and semantic content. It generates intermediate hidden states and predicts semantic tokens.

The Slow Transformer functions at an elevated level of abstraction, processing input text embeddings to encode global linguistic structures and semantic content. This module is responsible for generating intermediate hidden states and predicting semantic tokens with high precision.

Given an input sequence of tokens $\mathbf{x} = [x_1, x_2, \dots, x_T]$, the Slow Transformer generates hidden states $\mathbf{h} \in \mathbb{R}^{T \times D}$ and token logits $\mathbf{z}$ through the following transformations:
\begin{align}
    \mathbf{h} &= \text{SlowTransformer}(\mathbf{x}) \\
    \mathbf{z} &= \mathbf{W}_\text{tok} \cdot \text{Norm}(\mathbf{h})
\end{align}

where $\text{Norm}(\cdot)$ represents layer normalization, and $\mathbf{W}_\text{tok}$ denotes the learnable parameters of the token prediction layer.

\textbf{\subsubsubsection{Fast Transformer}}

The Fast Transformer refines the Slow Transformer's output through codebook embedding processing, capturing detailed acoustic features needed for natural speech. It processes residual information and optimizes codebook usage.

The Fast Transformer takes as input the concatenated sequence of hidden states $\mathbf{h}$ and codebook embeddings $\mathbf{c}$ according to:
\begin{equation}
    \mathbf{\tilde{h}} = [\mathbf{h}; \mathbf{c}],(\mathbf{h}^\text{fast})
\end{equation}
\begin{equation}
    \mathbf{h}^\text{fast} = \text{FastTransformer}(\mathbf{\tilde{h}},(\mathbf{h}^\text{fast}))
\end{equation}
\begin{equation}
    \mathbf{y} = \mathbf{W}_\text{cbk} \cdot \text{Norm}(\mathbf{h}^\text{fast})
\end{equation}

where $[\mathbf{h}; \mathbf{c}]$ represents the concatenation operation of $\mathbf{h}$ and $\mathbf{c}$, $\mathbf{W}_\text{cbk}$ comprises the learnable parameters of the codebook prediction layer, and $\mathbf{y} $ denotes the resultant codebook logits.

\subsubsection{Advantages of the Dual-AR Architecture}

The Dual-AR architecture in Fish-Speech demonstrates several significant advantages:
\begin{enumerate}
    \item \textbf{Enhanced Sequence Generation Stability}: The hierarchical processing of global and local information significantly improves the stability of GFSQ in sequence generation tasks.
    \item \textbf{Optimized Codebook Processing}: The Fast Transformer implements an efficient mechanism for codebook embedding processing that achieves improved performance without significant computational overhead, particularly for models of scale 7B or larger.
    \item \textbf{Superior Speech Synthesis Quality}: The synergistic interaction between Slow and Fast Transformers enables high-fidelity speech synthesis capable of handling complex linguistic phenomena.
    \item \textbf{Advanced Multilingual Processing}: The integration with Large Language Models (LLMs) for linguistic feature generation eliminates traditional grapheme-to-phoneme conversion dependencies, thereby streamlining the synthesis pipeline and enhancing multilingual capabilities. By mixing text data the comprehension will be further enhanced.
\end{enumerate}

\subsection{Firefly-GAN}
Firefly-GAN (FF-GAN) is an enhanced version of the EVA-GAN architecture with significant structural improvements. It replaces the traditional convolutional components of HiFi-GAN [\cite{kong2020hifi}] with a more efficient design, featuring a ParallelBlock instead of the Multi-Receptive Field (MRF) module. By incorporating Grouped Finite Scalar Vector Quantization (GFSQ), FF-GAN achieves better sequence generation stability and improved handling of linguistic variations, making it particularly effective for multilingual synthesis in AI applications.

\begin{figure}[h]
\vspace{-2pt}
\centering
\includegraphics[width=0.75\textwidth]{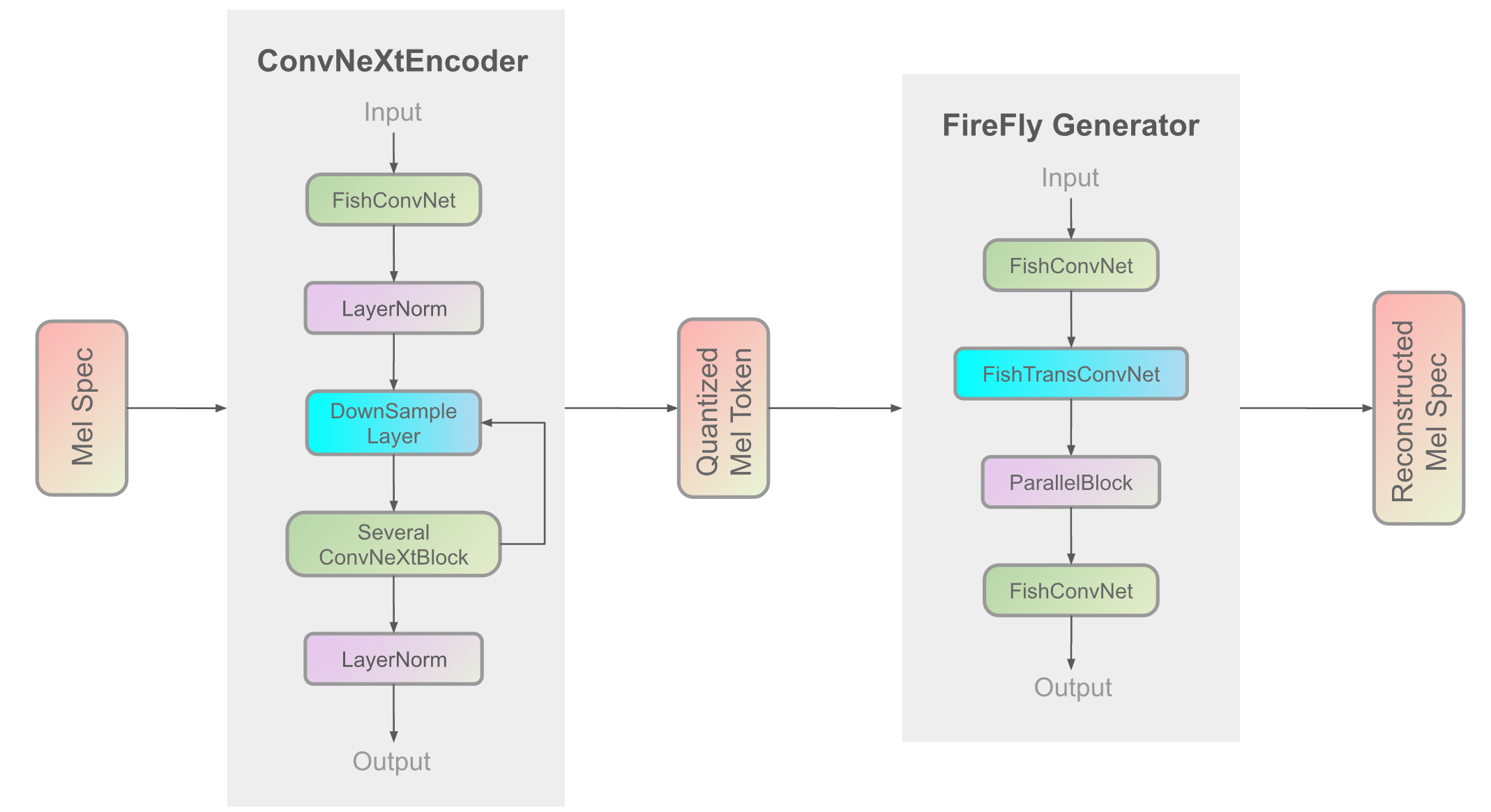}
\caption{FireFly GAN Architecture}
\label{fig:firefly-gan-arch}
\end{figure}

\subsubsection{Firefly Generator}
FF-GAN use enhanced Conv stucture including depth-wise separable convolution \cite{howard2017mobilenets} and  dilated convolutions \cite{yu2015multi}, replacing conventional Conv1d layers. This architectural refinement enhances the model's capacity to capture and synthesize complex audio features.

In our architecture, the conventional Multi-Receptive Field (MRF) module is replaced by a ParallelBlock, optimizing typo-codebook input processing efficiency. The ParallelBlock implements configurable convolution kernel sizes and dilation rates, utilizing stack-and-average mechanism for processing outputs from three ResBlocks, instead of directly addition operations. ParallelBlock \cite{liao2024eva} offers enhanced receptive field coverage, superior feature extraction capabilities, and improved configurability, contributing to higher-quality audio synthesis.

\subsubsection{Quantization Techniques}
In order to fit the typo-codebook task, we use Grouped Finite Scalar Vector Quantization (GFSQ) to act as a vq codebook in our system. The text below will elaborately tell you how we develop the GFSQ.

Given an input tensor $\mathbf{z} \in \mathbb{R}^{B \times C \times L}$. The entire process includes the following steps:
\textbf{\subsubsubsection{Downsampling}}

Using a downsampling function $f_{\text{down}}$ to downsample the input tensor $\mathbf{z}$, resulting in a downsampled tensor $\mathbf{z}_d \in \mathbb{R}^{B \times C_d \times L_d}$:
\begin{equation}
    \mathbf{z}_d = f_{\text{down}}(\mathbf{z})
\end{equation}
\textbf{\subsubsubsection{GFSQ Process}}
\begin{itemize}
    \item \textbf{Feature Grouping}
    Input feature matrix $\mathbf{Z}$ is divided into $G$ groups:
    \begin{equation}
        \mathbf{Z} = [\mathbf{Z}^{(1)}, \mathbf{Z}^{(2)}, \dots, \mathbf{Z}^{(G)}]
    \end{equation}

    \item \textbf{Scalar Quantization}
    For each scalar $z_{b,c,l}^{(g)}$:
    \begin{equation}
        \hat{z}_{b,c,l}^{(g)} = Q(z_{b,c,l}^{(g)})
    \end{equation}

    \item \textbf{Index Generation}
    Each scalar maps to index $k_{b,c,l}^{(g)}$

    \item \textbf{Decoding}
    \begin{equation}
        \hat{z}_{b,c,l}^{(g)} = \text{Codebook}^{(g)}[k_{b,c,l}^{(g)}]
    \end{equation}
\end{itemize}
\textbf{\subsubsubsection{Reconstruct the Quantized Downsampled Tensor}}
    
Concatenate the quantized vectors of all groups along the channel dimension to obtain the quantized downsampled tensor $\mathbf{z}_{q_d} \in \mathbb{R}^{B \times C_d \times L_d}$:
    \begin{equation}
        \mathbf{z}_{q_d}(b, :, l) = \left[ \mathbf{z}_{q_d}^{(1)}(b, :, l); \mathbf{z}_{q_d}^{(2)}(b, :, l); \dots; \mathbf{z}_{q_d}^{(G)}(b, :, l) \right]
    \end{equation}
\textbf{\subsubsubsection{Upsampling}}
    
Use the upsampling function $f_{\text{up}}$ to restore the quantized downsampled tensor to its original size, resulting in the final quantized tensor $\mathbf{z}_q \in \mathbb{R}^{B \times C \times L}$:
    \begin{equation}
        \mathbf{z}_q = f_{\text{up}}(\mathbf{z}_{q_d})
    \end{equation}

The goal is to make $\mathbf{z}_q$ approximate the original input $\mathbf{z}$ as closely as possible:
    \begin{equation}
        \mathbf{z}_q \approx \mathbf{z}
    \end{equation}
    
\subsubsection{Conclusion}

Our implementation of using GFSQ techniques achieves nearly 100\% codebook utilization, and gain better objetive and subjective score in our inner ablation than other quantization techiques like RFSQ, RVQ and GRFSQ. FF-GAN significantly enhancing stability in typo-codebook operations and ensuring comprehensive intermediate variable information retention in multi-emotional and multilingual tasks.

FF-GAN's innovative approach to typo-codebook stability has being already used in various song and music generation applications. The framework's performance and architectural could make it as a reference model for future AI agent development.

\section{Training and Inference}
\subsection{Training}
Fish Speech uses a three-stage training approach: initial pre-training with large batches of standard data, followed by SFT using smaller batches of high-quality data, and finally DPO training using manually labeled positive and negative sample pairs.

The training infrastructure was split into two components Fig.~\ref{fig:fs_architecture}: The AR training utilized an 8*H100 80G GPUs for one week, while the vocoder training employed an 8*4090 GPUs for an additional week. Note that these timelines exclude the DPO stage.

\subsection{Inference}
Our inference strategy follows the architecture in Fig.~\ref{fig:fs_architecture}. Using fish-tech including KV-cache [\cite{pope2023efficiently}], torch compile and other acceleration methodologies, the system achieves real-time factors of approximately 1:5 on consumer-grade NVIDIA RTX 4060 mobile platforms and 1:15 on high-performance NVIDIA RTX 4090 configurations. These architectural optimizations significantly improve latency in inferring, achieving a first-packet latency of 150ms.

Furthermore, the system can process information in flow, making it easy to work with modern AI tools and use them in different situations.

\section{Dataset}
Our training data includes a large collection of speech samples from both public sources and our own data collection process. The dataset contains about 720,000 hours of speech across different languages, with 300,000 hours each of English and Mandarin Chinese as the main components. We also included 20,000 hours each of other language families: Germanic (German), Romance (French, Italian), East Asian (Japanese, Korean), and Semitic (Arabic).

We carefully balanced the data across languages to help the model learn multiple languages at once. This approach helps the model perform well  generating mixed-language content. The large size and variety of our dataset significantly improves the model's ability to handle multiple languages naturally.

\section{Experimental Evaluation}\footnote{For experimental validation, we constrained our analysis to monolingual voice cloning scenarios, excluding cross-lingual synthesis tasks. The evaluation corpus comprised 10 distinct speakers (include different languages) identities, with 30 synthesized utterances per speaker, yielding a comprehensive evaluation set of 300 samples. It should be noted that cross-linguistic synthesis was not included in this evaluation.}
We conducted an evaluation for the speaker cloning task to access the effect of our architecture compared to the baseline models. The evaluation methodology includes both objective and subjective metrics: word error rate (WER) for intelligibility assessment, speaker embedding similarity measure for speech cloning fidelity assessment, and mean opinion score (MOS) for perceptual quality quantification. This evaluation framework aims to assess the model's ability to preserve speaker identity while maintaining high fidelity speech synthesis.

\subsection{Word Error Rate Analysis}
\begin{table}[htbp]
\centering
\begin{tabular}{l cc}
\toprule
\textbf{Model Name} & \textbf{WER(\%) \footnotemark} \\
\midrule
\textbf{Ground Truth} & \textbf{9.22} \\
\midrule
\textbf{fish-speech} & \textbf{6.89} \\
reecho & 11.92 \\
F5-TTS & 13.98 \\
CosyVoice & 22.20 \\
\bottomrule
\end{tabular}
\vspace{8pt}
\caption{Word Error Rate (WER) Results for Voice Cloning Tasks }
\label{tab:wer}
\end{table}
\footnotetext{Results obtained using OpenAI Whisper-medium ASR model for transcription evaluation.}

Analysis of Table \ref{tab:wer} shows that our model achieves a WER of 6.89\% in voice cloning tasks, which is not only much lower than the baseline models, but also exceeds the ground truth recordings (9.22\%). This performance provides strong evidence for the ability of our model in voice cloning scenarios. The gap between our model and competing models (ranging from 11.92\% to 22.20\%) underscores the improved synthetic stability and content fidelity of our methodology.

\subsection{Speaker Similarity Analysis}
\begin{table}[htbp]
\centering
\begin{tabular}{l cc}
\toprule
\textbf{Model Name} & \textbf{Resemblyzer} & \textbf{SpeechBrain\footnotemark} \\
\midrule
\textbf{Ground Truth} & \textbf{0.921} & \textbf{0.770} \\
\midrule
CosyVoice & 0.936 & 0.813 \\
\textbf{fish-speech} & \textbf{0.914} & \textbf{0.762} \\
F5-TTS & 0.905 & 0.787 \\
reecho & 0.887 & 0.636 \\
\bottomrule
\end{tabular}
\vspace{10pt}
\caption{Speaker similarity scores for different models, including ground truth.}
\label{tab:model_comparison}
\end{table}
\footnotetext{Experiments were conducted using the SpeechBrain framework~\cite{speechbrainV1}, version 1.0.1, as the underlying speech processing toolkit.}

Table \ref{tab:model_comparison} shows the effect of our typo-codebook strategy on speaker similarity metrics. Our fish-speech model achieves similarity scores of 0.914 and 0.762 on Resemblyzer and SpeechBrain, respectively, which are remarkably close to the ground truth performance (0.921 and 0.770). The gap of only 0.76\% from the ground truth in Resemblyzer and 1.04\% in SpeechBrain evaluations shows the superior ability of our model to capture natural speech characteristics. The results strongly suggest that our typo-codebook architecture enables a more comprehensive capture of acoustic states, leading to improved timbral fidelity of synthesized speech. Our approach significantly outperforms baseline models such as F5-TTS (0.905 and 0.787) and reecho (0.887 and 0.636). The consistent performance in both evaluation frameworks proves the validity of our method in preserving speaker features, which is crucial for high-quality text-to-speech synthesis and agent tasks.

\subsection{Perceptual Quality Assessment}
\begin{table}[htbp]
\centering
\begin{tabular}{l cc}
\toprule
\textbf{Model Name} & \textbf{MOS} \\
\midrule
\textbf{Ground Truth} & \textbf{5.00} \\
\midrule
\textbf{fish-speech} & \textbf{4.05} \\
CosyVoice & 3.80 \\
F5-TTS & 2.90 \\
reecho & 3.76 \\
\bottomrule
\end{tabular}
\vspace{10pt}
\caption{Five-scale Mean Opinion Score (MOS) Ratings of Cloned Voice Quality}
\label{tab:mos}
\end{table}

To evaluate the perceptual quality of synthesized audio, we conducted a comprehensive Mean Opinion Score (MOS) listening test with naive listeners who had no prior experience with audio processing. The evaluation followed a double-blind, randomized method to ensure an unbiased evaluation. The results show that fish-speech achieved significantly higher subjective scores compared to other baseline models (p < 0.05), demonstrating superior performance in terms of speech naturalness and speaker similarity. This evaluation in human perception metrics strongly suggests that fish-speech can better capture and reproduce the natural characteristics of human speech, especially in the context of voice cloning tasks.

\section{Conclusion}
Our research represents a significant advance in the field of text-to-speech (TTS) by introducing a novel multilingual and multi-emotional stabilization solution. The core innovation lies in our development of a typo-codebook vocoder integrated with a dual autoregressive (dual-AR) generation architecture. This architectural combination shows stability in the synthesis process while preserving acoustic features within the generated speech. Furthermore, our work utilizes a non-grapheme-to-phoneme (non-G2P) structure, an approach that effectively addresses limitations inherent in traditional phoneme-based systems and provides a robust foundation for cross-lingual and emotionally diverse TTS applications, particularly in the context of AI agent interactions.

\section{Future Work}
Building on these foundations, we propose several directions for future research. We plan to improve the performance of our model by integrating reinforcement learning techniques, focusing on improving cross-lingual generalization and emotional stability. We are also developing the Fish Agent application, an end-to-end language model based on our Fish-Speech framework. A preliminary demonstration of this system is currently available at \href{https://fish.audio/demo/live}{fish.audio/demo/live}. We remain committed to the open source community and will continue to maintain and extend our codebase to provide broader access to these technologies for researchers and developers.br

\bibliographystyle{unsrtnat}
\bibliography{references}  

\begin{thebibliography}{37}
\providecommand{\natexlab}[1]{#1}
\providecommand{\url}[1]{\texttt{#1}}
\expandafter\ifx\csname urlstyle\endcsname\relax
  \providecommand{\doi}[1]{doi: #1}\else
  \providecommand{\doi}{doi: \begingroup \urlstyle{rm}\Url}\fi

\bibitem[Wang et~al.(2023)Wang, Chen, Wu, Zhang, Zhou, Liu, Chen, Liu, Wang, Li, et~al.]{wang2023neural}
Chengyi Wang, Sanyuan Chen, Yu~Wu, Ziqiang Zhang, Long Zhou, Shujie Liu, Zhuo Chen, Yanqing Liu, Huaming Wang, Jinyu Li, et~al.
\newblock Neural codec language models are zero-shot text to speech synthesizers.
\newblock \emph{arXiv preprint arXiv:2301.02111}, 2023.

\bibitem[Kim et~al.(2021)Kim, Kong, and Son]{kim2021conditional}
Jaehyeon Kim, Jungil Kong, and Juhee Son.
\newblock Conditional variational autoencoder with adversarial learning for end-to-end text-to-speech.
\newblock In \emph{International Conference on Machine Learning}, pages 5530--5540. PMLR, 2021.

\bibitem[Ren et~al.(2020)Ren, Hu, Tan, Qin, Zhao, Zhao, and Liu]{ren2020fastspeech}
Yi~Ren, Chenxu Hu, Xu~Tan, Tao Qin, Sheng Zhao, Zhou Zhao, and Tie-Yan Liu.
\newblock Fastspeech 2: Fast and high-quality end-to-end text to speech.
\newblock \emph{arXiv preprint arXiv:2006.04558}, 2020.

\bibitem[Klatt(1987)]{klatt1987review}
Dennis~H Klatt.
\newblock Review of text-to-speech conversion for english.
\newblock \emph{The Journal of the Acoustical Society of America}, 82\penalty0 (3):\penalty0 737--793, 1987.

\bibitem[Casanova et~al.(2022)Casanova, Weber, Shulby, Junior, G{\"o}lge, and Ponti]{casanova2022yourtts}
Edresson Casanova, Julian Weber, Christopher~D Shulby, Arnaldo~Candido Junior, Eren G{\"o}lge, and Moacir~A Ponti.
\newblock Yourtts: Towards zero-shot multi-speaker tts and zero-shot voice conversion for everyone.
\newblock In \emph{International Conference on Machine Learning}, pages 2709--2720. PMLR, 2022.

\bibitem[Yang et~al.(2023)Yang, Tian, Tan, Huang, Liu, Chang, Shi, Zhao, Bian, Wu, et~al.]{yang2023uniaudio}
Dongchao Yang, Jinchuan Tian, Xu~Tan, Rongjie Huang, Songxiang Liu, Xuankai Chang, Jiatong Shi, Sheng Zhao, Jiang Bian, Xixin Wu, et~al.
\newblock Uniaudio: An audio foundation model toward universal audio generation.
\newblock \emph{arXiv preprint arXiv:2310.00704}, 2023.

\bibitem[Du et~al.(2024)Du, Chen, Zhang, Hu, Lu, Yang, Hu, Zheng, Gu, Ma, et~al.]{du2024cosyvoice}
Zhihao Du, Qian Chen, Shiliang Zhang, Kai Hu, Heng Lu, Yexin Yang, Hangrui Hu, Siqi Zheng, Yue Gu, Ziyang Ma, et~al.
\newblock Cosyvoice: A scalable multilingual zero-shot text-to-speech synthesizer based on supervised semantic tokens.
\newblock \emph{arXiv preprint arXiv:2407.05407}, 2024.

\bibitem[Mehta et~al.(2024)Mehta, Tu, Beskow, Sz{\'e}kely, and Henter]{mehta2024matcha}
Shivam Mehta, Ruibo Tu, Jonas Beskow, {\'E}va Sz{\'e}kely, and Gustav~Eje Henter.
\newblock Matcha-tts: A fast tts architecture with conditional flow matching.
\newblock In \emph{ICASSP 2024-2024 IEEE International Conference on Acoustics, Speech and Signal Processing (ICASSP)}, pages 11341--11345. IEEE, 2024.

\bibitem[Betker(2023)]{betker2023better}
James Betker.
\newblock Better speech synthesis through scaling.
\newblock \emph{arXiv preprint arXiv:2305.07243}, 2023.

\bibitem[Mentzer et~al.(2023)Mentzer, Minnen, Agustsson, and Tschannen]{mentzer2023finite}
Fabian Mentzer, David Minnen, Eirikur Agustsson, and Michael Tschannen.
\newblock Finite scalar quantization: Vq-vae made simple.
\newblock \emph{arXiv preprint arXiv:2309.15505}, 2023.

\bibitem[Tan et~al.(2021)Tan, Qin, Soong, and Liu]{tan2021survey}
Xu~Tan, Tao Qin, Frank Soong, and Tie-Yan Liu.
\newblock A survey on neural speech synthesis.
\newblock \emph{arXiv preprint arXiv:2106.15561}, 2021.

\bibitem[Ren et~al.(2019)Ren, Tan, Qin, Zhao, Zhao, and Liu]{ren2019almost}
Yi~Ren, Xu~Tan, Tao Qin, Sheng Zhao, Zhou Zhao, and Tie-Yan Liu.
\newblock Almost unsupervised text to speech and automatic speech recognition.
\newblock In \emph{International conference on machine learning}, pages 5410--5419. PMLR, 2019.

\bibitem[Capes et~al.(2017)Capes, Coles, Conkie, Golipour, Hadjitarkhani, Hu, Huddleston, Hunt, Li, Neeracher, et~al.]{capes2017siri}
Tim Capes, Paul Coles, Alistair Conkie, Ladan Golipour, Abie Hadjitarkhani, Qiong Hu, Nancy Huddleston, Melvyn Hunt, Jiangchuan Li, Matthias Neeracher, et~al.
\newblock Siri on-device deep learning-guided unit selection text-to-speech system.
\newblock In \emph{Interspeech}, pages 4011--4015, 2017.

\bibitem[Van Den~Oord et~al.(2016)Van Den~Oord, Dieleman, Zen, Simonyan, Vinyals, Graves, Kalchbrenner, Senior, Kavukcuoglu, et~al.]{van2016wavenet}
Aaron Van Den~Oord, Sander Dieleman, Heiga Zen, Karen Simonyan, Oriol Vinyals, Alex Graves, Nal Kalchbrenner, Andrew Senior, Koray Kavukcuoglu, et~al.
\newblock Wavenet: A generative model for raw audio.
\newblock \emph{arXiv preprint arXiv:1609.03499}, 12, 2016.

\bibitem[Kalchbrenner et~al.(2018)Kalchbrenner, Elsen, Simonyan, Noury, Casagrande, Lockhart, Stimberg, Oord, Dieleman, and Kavukcuoglu]{kalchbrenner2018efficient}
Nal Kalchbrenner, Erich Elsen, Karen Simonyan, Seb Noury, Norman Casagrande, Edward Lockhart, Florian Stimberg, Aaron Oord, Sander Dieleman, and Koray Kavukcuoglu.
\newblock Efficient neural audio synthesis.
\newblock In \emph{International Conference on Machine Learning}, pages 2410--2419. PMLR, 2018.

\bibitem[Chen et~al.(2020)Chen, Zhang, Zen, Weiss, Norouzi, and Chan]{chen2020wavegrad}
Nanxin Chen, Yu~Zhang, Heiga Zen, Ron~J Weiss, Mohammad Norouzi, and William Chan.
\newblock Wavegrad: Estimating gradients for waveform generation.
\newblock \emph{arXiv preprint arXiv:2009.00713}, 2020.

\bibitem[Kong et~al.(2020)Kong, Kim, and Bae]{kong2020hifi}
Jungil Kong, Jaehyeon Kim, and Jaekyoung Bae.
\newblock Hifi-gan: Generative adversarial networks for efficient and high fidelity speech synthesis.
\newblock \emph{Advances in neural information processing systems}, 33:\penalty0 17022--17033, 2020.

\bibitem[Liao et~al.(2024)Liao, Lan, and Zachariah]{liao2024eva}
Shijia Liao, Shiyi Lan, and Arun~George Zachariah.
\newblock Eva-gan: Enhanced various audio generation via scalable generative adversarial networks.
\newblock \emph{arXiv preprint arXiv:2402.00892}, 2024.

\bibitem[Van Den~Oord et~al.(2017)Van Den~Oord, Vinyals, et~al.]{van2017neural}
Aaron Van Den~Oord, Oriol Vinyals, et~al.
\newblock Neural discrete representation learning.
\newblock \emph{Advances in neural information processing systems}, 30, 2017.

\bibitem[Zeghidour et~al.(2021)Zeghidour, Luebs, Omran, Skoglund, and Tagliasacchi]{zeghidour2021soundstream}
Neil Zeghidour, Alejandro Luebs, Ahmed Omran, Jan Skoglund, and Marco Tagliasacchi.
\newblock Soundstream: An end-to-end neural audio codec.
\newblock \emph{IEEE/ACM Transactions on Audio, Speech, and Language Processing}, 30:\penalty0 495--507, 2021.

\bibitem[D{\'e}fossez et~al.(2022)D{\'e}fossez, Copet, Synnaeve, and Adi]{defossez2022high}
Alexandre D{\'e}fossez, Jade Copet, Gabriel Synnaeve, and Yossi Adi.
\newblock High fidelity neural audio compression.
\newblock \emph{arXiv preprint arXiv:2210.13438}, 2022.

\bibitem[Lacombe et~al.(2024)Lacombe, Srivastav, and Gandhi]{lacombe-etal-2024-parler-tts}
Yoach Lacombe, Vaibhav Srivastav, and Sanchit Gandhi.
\newblock Parler-tts.
\newblock \url{https://github.com/huggingface/parler-tts}, 2024.

\bibitem[Zhao et~al.(2023)Zhao, Yu, and Qin]{zhao2024melo}
Wenliang Zhao, Xumin Yu, and Zengyi Qin.
\newblock Melotts: High-quality multi-lingual multi-accent text-to-speech, 2023.
\newblock URL \url{https://github.com/myshell-ai/MeloTTS}.
\newblock Software.

\bibitem[Gao et~al.(2023)Gao, Morioka, Zhang, and Chen]{gao2023e3}
Yuan Gao, Nobuyuki Morioka, Yu~Zhang, and Nanxin Chen.
\newblock E3 tts: Easy end-to-end diffusion-based text to speech.
\newblock In \emph{2023 IEEE Automatic Speech Recognition and Understanding Workshop (ASRU)}, pages 1--8. IEEE, 2023.

\bibitem[Casanova et~al.(2024)Casanova, Davis, G{\"o}lge, G{\"o}knar, Gulea, Hart, Aljafari, Meyer, Morais, Olayemi, et~al.]{casanova2024xtts}
Edresson Casanova, Kelly Davis, Eren G{\"o}lge, G{\"o}rkem G{\"o}knar, Iulian Gulea, Logan Hart, Aya Aljafari, Joshua Meyer, Reuben Morais, Samuel Olayemi, et~al.
\newblock Xtts: a massively multilingual zero-shot text-to-speech model.
\newblock \emph{arXiv preprint arXiv:2406.04904}, 2024.

\bibitem[Liu and Mak(2019)]{liu2019cross}
Zhaoyu Liu and Brian Mak.
\newblock Cross-lingual multi-speaker text-to-speech synthesis for voice cloning without using parallel corpus for unseen speakers.
\newblock \emph{arXiv preprint arXiv:1911.11601}, 2019.

\bibitem[Nekvinda and Du{\v{s}}ek(2020)]{nekvinda2020one}
Tom{\'a}{\v{s}} Nekvinda and Ond{\v{r}}ej Du{\v{s}}ek.
\newblock One model, many languages: Meta-learning for multilingual text-to-speech.
\newblock \emph{arXiv preprint arXiv:2008.00768}, 2020.

\bibitem[Li et~al.(2019)Li, Zhang, Sainath, Wu, and Chan]{li2019bytes}
Bo~Li, Yu~Zhang, Tara Sainath, Yonghui Wu, and William Chan.
\newblock Bytes are all you need: End-to-end multilingual speech recognition and synthesis with bytes.
\newblock In \emph{ICASSP 2019-2019 IEEE International Conference on Acoustics, Speech and Signal Processing (ICASSP)}, pages 5621--5625. IEEE, 2019.

\bibitem[Kumar et~al.(2024)Kumar, Seetharaman, Luebs, Kumar, and Kumar]{kumar2024high}
Rithesh Kumar, Prem Seetharaman, Alejandro Luebs, Ishaan Kumar, and Kundan Kumar.
\newblock High-fidelity audio compression with improved rvqgan.
\newblock \emph{Advances in Neural Information Processing Systems}, 36, 2024.

\bibitem[Chen et~al.(2023)Chen, Watanabe, and Rudnicky]{chen2023vector}
Li-Wei Chen, Shinji Watanabe, and Alexander Rudnicky.
\newblock A vector quantized approach for text to speech synthesis on real-world spontaneous speech.
\newblock In \emph{Proceedings of the AAAI Conference on Artificial Intelligence}, volume~37, pages 12644--12652, 2023.

\bibitem[Wang et~al.(2019)Wang, Takaki, Yamagishi, King, and Tokuda]{wang2019vector}
Xin Wang, Shinji Takaki, Junichi Yamagishi, Simon King, and Keiichi Tokuda.
\newblock A vector quantized variational autoencoder (vq-vae) autoregressive neural $ f\_0 $ model for statistical parametric speech synthesis.
\newblock \emph{IEEE/ACM Transactions on Audio, Speech, and Language Processing}, 28:\penalty0 157--170, 2019.

\bibitem[Vaswani(2017)]{vaswani2017attention}
A~Vaswani.
\newblock Attention is all you need.
\newblock \emph{Advances in Neural Information Processing Systems}, 2017.

\bibitem[Subakan et~al.(2021)Subakan, Ravanelli, Cornell, Bronzi, and Zhong]{subakan2021attention}
Cem Subakan, Mirco Ravanelli, Samuele Cornell, Mirko Bronzi, and Jianyuan Zhong.
\newblock Attention is all you need in speech separation.
\newblock In \emph{ICASSP 2021-2021 IEEE International Conference on Acoustics, Speech and Signal Processing (ICASSP)}, pages 21--25. IEEE, 2021.

\bibitem[Howard(2017)]{howard2017mobilenets}
Andrew~G Howard.
\newblock Mobilenets: Efficient convolutional neural networks for mobile vision applications.
\newblock \emph{arXiv preprint arXiv:1704.04861}, 2017.

\bibitem[Yu(2015)]{yu2015multi}
F~Yu.
\newblock Multi-scale context aggregation by dilated convolutions.
\newblock \emph{arXiv preprint arXiv:1511.07122}, 2015.

\bibitem[Pope et~al.(2023)Pope, Douglas, Chowdhery, Devlin, Bradbury, Heek, Xiao, Agrawal, and Dean]{pope2023efficiently}
Reiner Pope, Sholto Douglas, Aakanksha Chowdhery, Jacob Devlin, James Bradbury, Jonathan Heek, Kefan Xiao, Shivani Agrawal, and Jeff Dean.
\newblock Efficiently scaling transformer inference.
\newblock \emph{Proceedings of Machine Learning and Systems}, 5:\penalty0 606--624, 2023.

\bibitem[Ravanelli et~al.(2024)Ravanelli, Parcollet, Moumen, de~Langen, Subakan, Plantinga, Wang, Mousavi, Libera, Ploujnikov, Paissan, Borra, Zaiem, Zhao, Zhang, Karakasidis, Yeh, Champion, Rouhe, Braun, Mai, Zuluaga-Gomez, Mousavi, Nautsch, Liu, Sagar, Duret, Mdhaffar, Laperriere, Rouvier, Mori, and Esteve]{speechbrainV1}
Mirco Ravanelli, Titouan Parcollet, Adel Moumen, Sylvain de~Langen, Cem Subakan, Peter Plantinga, Yingzhi Wang, Pooneh Mousavi, Luca~Della Libera, Artem Ploujnikov, Francesco Paissan, Davide Borra, Salah Zaiem, Zeyu Zhao, Shucong Zhang, Georgios Karakasidis, Sung-Lin Yeh, Pierre Champion, Aku Rouhe, Rudolf Braun, Florian Mai, Juan Zuluaga-Gomez, Seyed~Mahed Mousavi, Andreas Nautsch, Xuechen Liu, Sangeet Sagar, Jarod Duret, Salima Mdhaffar, Gaelle Laperriere, Mickael Rouvier, Renato~De Mori, and Yannick Esteve.
\newblock Open-source conversational ai with {SpeechBrain} 1.0, 2024.
\newblock URL \url{https://arxiv.org/abs/2407.00463}.

\end{thebibliography}

\appendix
\section{Training Details}
We trained our model on NVIDIA H100 cluster with the following hyperparameters:

\paragraph{Optimization:}
\begin{itemize}
    \item Optimizer: AdamW ($\beta_1=0.9$, $\beta_2=0.98$, $\epsilon=10^{-8}$)
    \item LR: $5\times10^{-4}$ 
    \item Weight decay: 0.01
\end{itemize}

\paragraph{Training Config:}
\begin{itemize}
    \item Batch size: 1M tokens
    \item Training steps: 500K
    \item LR schedule: Cosine decay with warmup
    \item Warmup steps: 2K
    \item Final LR ratio: 0.1
\end{itemize}
\end{document}